\documentclass[aip,jcp,reprint,superscriptaddress,twocolumn,longbibliography,floatfix,eqsecnum]{revtex4-2}

\usepackage{amsmath,amssymb,graphicx}
\usepackage{mathpazo}
\usepackage{bm}
\usepackage[unicode=true,
bookmarks=true,bookmarksnumbered=false,bookmarksopen=false,
breaklinks=false,pdfborder={0 0 0},pdfborderstyle={},backref=false,colorlinks=true]
{hyperref}
\hypersetup{pdfborderstyle={},pdfborderstyle={},pdfborderstyle={},pdfborderstyle={},pdfborderstyle={},pdfborderstyle={},linkcolor=blue,citecolor=blue}
\usepackage{breakurl}

\usepackage{natbib}

\newcommand\beq{\begin{equation}}
\newcommand\eeq{\end{equation}}
\newcommand\beqa{\begin{eqnarray}}
\newcommand\eeqa{\end{eqnarray}}
\newcommand{\nn}{\nonumber\\}
\def\bal#1\eal{\begin{align}#1\end{align}}

\newcommand{\bp}{\beta p}

\newcommand{\pt}{\gamma}

\usepackage[dvipsnames]{xcolor}
\usepackage[normalem]{ulem}

\begin{document}
	
	\title{Thermodynamic and structural behavior of one-dimensional divalent patchy hard rods: Wertheim's first-order thermodynamic perturbation theory vs exact results}
	
	\author{Ana M. Montero}
	\email{anamontero@unex.es}
	\affiliation{Departamento de F\'isica,
		Universidad de Extremadura, E-06006 Badajoz, Spain}
	\author{Andr\'es Santos}
	\email{andres@unex.es}
	\affiliation{Departamento de F\'isica,
		Universidad de Extremadura, E-06006 Badajoz, Spain}
	\affiliation{Instituto de Computaci\'on Cient\'ifica Avanzada (ICCAEx),
		Universidad de Extremadura, E-06006 Badajoz, Spain}
	\author{P\'eter Gurin}
	\email{gurin.peter@mk.uni-pannon.hu}
	\affiliation{Physics Department, Centre for Natural Sciences,
		University of Pannonia, P.O. Box 158, Veszpr\'em H-8201, Hungary}
	\author{Szabolcs Varga}
	\email{varga.szabolcs@mk.uni-pannon.hu}
	\affiliation{Physics Department, Centre for Natural Sciences,
		University of Pannonia, P.O. Box 158, Veszpr\'em H-8201, Hungary}

	\begin{abstract}
          We investigate the thermodynamic and structural properties of divalent patchy hard rods confined to a one-dimensional channel by modeling the bonding sites as attractive square-well (SW) patches located at the rod tips. The zero-range sticky limit is recovered by letting the well width vanish while keeping the stickiness parameter finite. While Wertheim's first-order thermodynamic perturbation theory becomes exact in this sticky limit, it fails for finite-range site--site interactions.
         Because the present model is mathematically equivalent to an exactly solvable one-dimensional nearest-neighbor fluid, we use the exact solution to reformulate the thermodynamics in terms of association-theory variables, including the fraction of unbonded sites and a generalized law of mass action.
          Finite-range SW sites produce a richer structural behavior than sticky sites, including monotonic and oscillatory asymptotic decay of the pair correlation function, separated by the Fisher--Widom line. In the monotonic regime, the correlation length exhibits an absolute maximum defining the Widom line, while in the oscillatory regime it may display a local maximum and minimum, whose locus defines the ``Extrema of the Correlation length under Oscillatory decay'' line. These features disappear in the sticky limit, where the system remains entirely in the oscillatory regime. We also show that the high-pressure behavior of the correlation length changes from $\xi\sim p^2$ for finite-range SW sites to $\xi\sim p^3$ in the sticky limit.
	\end{abstract}
	
	\maketitle
	
	\section{Introduction}\label{sec:introduction}

In low-valence patchy colloidal systems, restricted coordination numbers dictate the emergence of distinct hierarchical architectures.\cite{BBL11,RS11c,BCCRV17} These morphologies include a wide variety of structures, ranging from flexible linear chains and branched clusters to rings and intricate network topologies.\cite{SPY13,BTLS23} Although the constituent building blocks are relatively simple, the resulting collective behavior is governed by a subtle interplay of directional interactions arising from the spatial arrangement and range of patches on the particle surface.\cite{MK99,MHSRM19}
Consequently, a rigorous theoretical description is required to understand
how the geometry and range of directional interactions determine the
formation of complex self-assembled structures.\cite{ZG04,DBK10,DL21,RLMS22}

Divalent patchy particles constitute a fundamental class of building blocks for the self-assembly of low-valence colloidal and polymeric systems.\cite{LZLQS14,SXSLCSG19,SRJRGSBBS21} Characterized by two bonding sites, they provide a direct link to systems of colloids or polymers with reactive functional groups located at both ends.\cite{MBS12,MCZA16,NDM18,JM21} The resulting structures are polydisperse ensembles of linear chains and rings, whose properties are governed by the equilibrium between bond formation and breakage. Therefore, these systems occupy an intermediate regime between simple monomeric fluids and highly connected network-forming systems. Their relative simplicity, combined with nontrivial self-assembly, makes them ideal testbeds for assessing the predictive capabilities of association theories,\cite{JCG88,GVWMJB97,PGJ02,DG06,JSB22,A25} classical density functional theories,\cite{KR92,BP96,WL07,MC13,BGM24b} and computer simulation methods.\cite{MGTP95,MSJ97,LK04,SBDT07,S21}

Beyond bulk conditions, restricted geometries provide an additional source
of complexity and may qualitatively alter the equilibrium behavior of
associating fluids. Strong confinement, such as that imposed by narrow
channels or nanotube-like environments, can effectively reduce the accessible
dimensionality and give rise to phenomena absent in bulk systems.\cite{LP83,M16,M15c,KA16,GSPV26}

Although the present strictly one-dimensional (1D) model should primarily be
viewed as an exactly solvable theoretical framework rather than as a direct
representation of a specific experimental system, it  may serve as a useful reference for more realistic quasi-one-dimensional
fluids.\cite{T36,SZK53,VCR03,S16,MS19,FMS21,CR69,KP93,GV25}
More generally, it may also provide insight into reversible self-assembly processes dominated by strongly constrained linear aggregation, where chain branching and loop formation are largely suppressed.

It is customary to model patchy colloidal particles as rigid hard bodies---ranging from spheres to anisotropic rods---decorated with localized attractive sites on their surfaces.\cite{BBL11,RLMS22} These sites can be represented through different interaction potentials, including infinitesimal ``sticky'' spots, soft-core attractions, or finite-range wells, which effectively determine the particle valency.\cite{KF03,RMGLZ20} Such models enable a systematic investigation of the combined role of particle shape and directional attractions in determining collective behavior. In particular, hard bodies decorated with square-well (SW) sites have proven to be especially suitable for capturing the essential features of directional self-assembly in patchy systems.\cite{RRR18,KPHC24}
Recent developments have also explored the application of association theories to particles with anisotropic hard cores and more complex bonding constraints,\cite{K26} illustrating the continuing interest in extending Wertheim-type approaches beyond standard spherical reference systems.

The structural constraints imposed by single-file confinement, together with the representation of patchy particles as hard bodies with SW attractive sites, define an ideal framework for the application of Wertheim's first-order thermodynamic perturbation theory (TPT1).\cite{W84a,W84b,W86,W87} In strictly 1D systems, the topological assumptions underlying TPT1 are automatically satisfied: each site can form at most one bond, and both ring formation and multiple bonding between the same pair of particles are excluded by geometry. Since these limitations are a major source of inaccuracy in higher-dimensional applications, their absence in 1D suggests that TPT1 should attain its highest level of accuracy---and possibly exactness---under single-file confinement. Accordingly, 1D divalent patchy systems provide a stringent benchmark for assessing the performance of the theory in describing chain formation.
	
In this work, we pursue three main objectives. First, we assess the validity and limitations of Wertheim's TPT1 for divalent patchy hard rods with finite-range attractive sites by comparing its predictions with exact results. Second, we reformulate the exact solution in the language of association theories, deriving an exact generalized law of mass action and an exact bonding free-energy contribution. Third, we investigate the structural consequences of finite-range bonding and compare them with those of the zero-range sticky limit.

The resulting phenomenology turns out to be remarkably rich. At low and intermediate densities, attractive interactions dominate, whereas repulsive hard-core interactions control the behavior near close packing. We demonstrate that finite-range SW and zero-range sticky sites display qualitatively different behaviors. In particular: (i) the equation of state evolves from sticky-like to hard-rod-like behavior as pressure increases; (ii) the Fisher--Widom (FW) and Widom lines emerge only for finite-range interactions; (iii) a novel structural line, hereafter referred to as the ``Extrema of the Correlation length under Oscillatory decay'' (ECO) line, appears in the oscillatory regime; and (iv) the asymptotic divergence of the correlation length changes from $\xi\sim p^2$ for finite-range interactions to $\xi\sim p^3$ in the sticky limit.

We further show that standard TPT1 is exact only in the zero-range sticky limit, while finite-range attractions require modifications of both the law of mass action and the bonding free-energy contribution. By exploiting the exact solution of the 1D nearest-neighbor problem, we
recast it in the language of association theories, thereby obtaining exact expressions for bonding quantities and highlighting the limitations of standard TPT1.

The remainder of the paper is organized as follows. Section~\ref{sec:system} introduces the zero-range sticky and finite-range SW models used to describe divalent patchy hard rods. Section~\ref{sec:Wertheim_theory} reviews Wertheim's TPT1 framework. The exact solution is presented in Sec.~\ref{sec:ND_theory}.
Section~\ref{sec:SW_SHS} discusses the thermodynamic and structural consequences of finite-range bonding, including the equation of state, chain statistics, correlation length, and the associated structural crossover lines.
Finally, Sec.~\ref{sec:conclusions} summarizes the main findings and outlines possible directions for future work. The additional technical details are provided in the Appendixes.

	\section{Hard rods decorated with SW sites}\label{sec:system}
	
\begin{figure}
	\includegraphics[width=\linewidth]{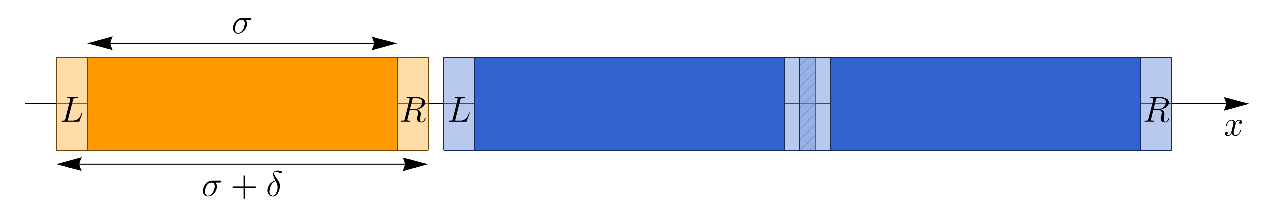}
	\caption{Schematic representation of divalent patchy hard rods of length $\sigma$ aligned along the $x$ axis. Particles cannot overlap (i.e.,  the distance between the centers of two neighboring rods must be $x>\sigma$), but neighboring rods can form bonds when $\sigma<x<\sigma+\delta$ through the interaction between the right ($R$) site of the left particle and the left ($L$) site of the right particle. A monomer and a dimer are shown for illustration.}
	\label{fig:system}
\end{figure}

We study the thermodynamic behavior and structural properties of divalent patchy hard rods confined to a strictly 1D channel. In this geometry, particles cannot rotate, fluctuate in the transverse directions, or exchange their relative order along the channel. However, they are free to translate along a straight line (the $x$ axis), and each particle interacts only with its nearest neighbors to the left and to the right. As a consequence, there is a single bonding mechanism: the right ($R$) site of a given particle can bind to the left ($L$) site of its right neighbor. Bonds of type $R$--$R$ and $L$--$L$ are, therefore, excluded. A schematic representation of the system is shown in Fig.~\ref{fig:system}.

The interaction between neighboring particles is described by a SW hard-rod pair potential,
\begin{equation}
\label{eq:pairPotential}
u(x)=
\begin{cases}
\infty, & x<\sigma, \\
-\epsilon, & \sigma < x < \sigma+\delta, \\
0, & x>\sigma+\delta,
\end{cases}
\end{equation}
where $x$ is the center-to-center distance between adjacent particles, $\sigma$ is the rod length, $\epsilon$ is the well depth, and $\delta$ is the range of the attractive interaction between the $L$ and $R$ sites. To ensure that bonding occurs only between nearest neighbors (i.e., to exclude second-neighbor site--site interactions), the range of attraction is restricted to $0<\delta<\sigma$.

The present model is mathematically equivalent to the 1D SW fluid, for which an exact solution is available in the isothermal--isobaric ensemble.\cite{S16} The patchy interpretation adopted here provides a convenient framework for comparison with Wertheim's association theory.

The sticky limit of the interaction is recovered by taking $\delta\to 0$ and $\epsilon\to\infty$ while keeping the dimensionless combination
\beq
\label{stickiness}
\tau^{-1}=(e^{\beta\epsilon}-1)\frac{\delta}{\sigma}
\eeq
finite, where $\tau^{-1}$ defines the degree of ``stickiness.''

Since the aim of this work is to compare SW systems with different values of $\delta$, including the sticky limit, it is convenient to use the inverse stickiness parameter $\tau$ as the thermodynamic control variable instead of the temperature. In this way, systems with different well depths $\epsilon$ and ranges $\delta$ can be compared on an equal footing through the common reduced parameter $\tau$, while the sticky limit is naturally obtained by taking $\delta\to 0$ at fixed $\tau$.

	\section{Wertheim association theory}\label{sec:Wertheim_theory}
\subsection{General formulation of Wertheim's TPT1}
Consider $N$ divalent patchy rods confined to a line segment of length $L$. The Helmholtz free energy can be decomposed into ideal, excess (hard-body), and bonding contributions,
	\begin{equation}
\label{eq:freeEnergySplit}
{F} = {F_{\mathrm{id}}} + {F_{\mathrm{ex}}} + {F_{b}}.
\end{equation}
The ideal-gas contribution reads
	\begin{equation}
\label{eq:idealFreeEnergy}
\frac{\beta F_{\mathrm{id}}}{N} = \ln\rho  - 1,
\end{equation}
where $\beta=1/k_B T$ is the inverse temperature, $\rho=N/L$ is the number density, and the thermal de Broglie wavelength is set to unity. The excess free energy associated with the hard-core interaction is exactly given by	 \begin{equation}
\label{eq:excessFreeEnergy}
\frac{\beta F_{\mathrm{ex}}}{N} = -\ln(1-\rho \sigma).
\end{equation}

The bonding contribution is not known exactly in general. Within Wertheim's TPT1,\cite{W84a,W84b,W86,W87} it is assumed that each site can participate in at most one bond and that no more than one bond can be formed between a given pair of particles. In the present 1D system, these conditions are automatically satisfied. Within TPT1, the bonding free energy is expressed in terms of the fraction $X_i$ of sites of type $i$ that are not bonded,
\begin{equation}
\label{eq:bondFreeEnergyWertheim}
\frac{\beta F_b}{N} = \sum_i \left( \ln X_i - \frac{X_i}{2} + \frac{1}{2} \right),
\end{equation}
where the $X_i$ obey the law of mass action
\begin{equation}
\label{eq:lawMassActionGeneral}
\frac{1}{X_i} = 1 + \rho \sum_j X_j \Delta_{ij}.
\end{equation}
Here, $\Delta_{ij}$ denotes the bonding integral between sites $i$ and $j$.

\subsection{Application to 1D divalent patchy hard rods}

In the present model, $\Delta_{LL}=\Delta_{RR}=0$, while
$\Delta_{LR}=\Delta_{RL}=(e^{\beta\epsilon}-1)\int_{\sigma}^{\sigma+\delta}dx\, g_{\text{HR}}(x)$,
with $g_{\text{HR}}(x)=\exp[-\rho(x-\sigma)/(1-\rho\sigma)]/(1-\rho\sigma)$ the exact hard-rod pair distribution function for $\sigma<x\le\sigma+\delta$.\cite{SZK53} Carrying out the integral yields
\begin{equation}
\label{eq:bondingIntegral}
\Delta_{RL}=\left(e^{\beta\epsilon}-1\right)\frac{1-e^{-\rho\delta/(1-\rho\sigma)}}{\rho} .
\end{equation}
By symmetry, $X_L=X_R\equiv X$, and Eq.~\eqref{eq:lawMassActionGeneral} reduces to
\begin{equation}
\label{eq:lawMassActionFinite}
\frac{1}{X}=1+X\frac{1-e^{-\rho\delta/(1-\rho\sigma)}}{\tau\delta/\sigma} ,
\end{equation}
where we have used Eq.~\eqref{stickiness}. This law of mass action predicts a spurious nonzero fraction of unbonded sites in the close-packing limit,
\beq
\label{10}
\lim_{\rho\sigma\to 1} X=\frac{\tau\delta/\sigma}{2}\left(\sqrt{1+\frac{4}{\tau\delta/\sigma}}-1\right).
\eeq
Combining Eqs.~\eqref{eq:idealFreeEnergy}--\eqref{eq:bondFreeEnergyWertheim} with $X_L=X_R= X$, the total free energy per particle is
\begin{equation}
\label{eq:totalFreeEnergy}
\frac{\beta F}{N}=\ln \rho -1-\ln(1-\rho\sigma)+2\ln X-X+1 .
\end{equation}
Thus, the thermodynamics of the system is fully determined by Eq.~\eqref{eq:totalFreeEnergy} together with Eq.~\eqref{eq:lawMassActionFinite}. The explicit dependence on $\rho$ accounts for the hard-core contribution, while the dependence on $X$ encodes the effect of bonding, which vanishes in the absence of bonds ($X=1$).

The pressure follows from $\bp=\rho^2 \partial(\beta F/N)/\partial\rho$, yielding
\begin{equation}
\label{eq:pressureFinite}
\bp=\frac{\rho}{1-\rho\sigma}\left[1-\frac{\rho\sigma}{1-\rho\sigma}\frac{X^2}{\tau}e^{-\rho\delta/(1-\rho\sigma)}\right] ,
\end{equation}
where we used the identity $(2/X-1)\partial X/\partial\rho=-(X^2\sigma/\tau) \exp[-\rho\delta/(1-\rho\sigma)]/(1-\rho\sigma)^2$,
which follows from Eq.~\eqref{eq:lawMassActionFinite}.

In the high-temperature limit ($\tau\to\infty$), Eq.~\eqref{eq:pressureFinite} reduces to the Tonks equation of state for hard rods,\cite{T36}
\begin{equation}
\label{eq:tonksPressure}
\bp_T=\frac{\rho}{1-\rho\sigma} .
\end{equation}

In the sticky limit ($\delta\to0$ at fixed $\tau$), Eq.~\eqref{eq:lawMassActionFinite} becomes
\begin{equation}
\label{eq:lawMassActionSticky}
\frac{1}{X}=1+\frac{X}{\tau}\frac{\rho\sigma}{1-\rho\sigma} \quad \text{(sticky limit)},
\end{equation}
which is the law of mass action for sticky hard rods. In this case, $\lim_{\rho\sigma\to 1} X=0$. In addition in this limit, Eq.~\eqref{eq:pressureFinite} simplifies to
\begin{equation}
\label{eq:pressureSticky}
\bp=\frac{\rho}{1-\rho\sigma}X \quad \text{(sticky limit)},
\end{equation}
where Eq.~\eqref{eq:lawMassActionSticky} has been used. This expression shows that the ratio $p/p_T$ between the pressure of the sticky system and that of the reference hard-rod fluid is simply equal to the fraction of unbonded sites, $X$.
Note that combination of Eqs.~\eqref{eq:lawMassActionSticky} and \eqref{eq:pressureSticky} gives
\beq
\label{EOS-SHS}
\bp=\frac{\sqrt{1+4\tau^{-1}\frac{\rho\sigma}{1-\rho\sigma}}-1}{2\tau^{-1}\sigma} \quad \text{(sticky limit)}.
\eeq

\subsection{Cluster-size distribution and physical interpretation of $X$}

To elucidate the physical meaning of $X$, which plays a central role in Wertheim's association theory, consider the number density $\rho_i$ of chain molecules composed of $i$ rods. This quantity can be written as\cite{SBDT07}
\begin{equation}
\label{eq:clusterDensity}
\rho_i=\rho X^2(1-X)^{i-1} ,
\end{equation}
where the factor $(1-X)^{i-1}$ accounts for the $i-1$ bonds within the chain, while the prefactor $X^2$ reflects the two unbonded sites at the chain ends. Consistency with the total density follows from
\begin{equation}
\label{eq:densityNormalization}
\rho=\sum_{i=1}^{\infty} i\rho_i,
\end{equation}
together with the identity $\sum_{i=1}^{\infty} i(1-X)^{i-1}=1/X^2$.

The number density of chain molecules (including monomers), $\rho_M=N_M/L$, is given by $\rho_M=\sum_{i=1}^{\infty}\rho_i$. Using $\sum_{i=1}^{\infty}(1-X)^{i-1}=1/X$, one obtains
\begin{equation}
\label{eq:monomerDensity}
\rho_M=\rho X .
\end{equation}
Accordingly, the fraction of $i$-mers reads
\begin{equation}
\label{eq:clusterSizeDistribution}
f_i=\frac{\rho_i}{\rho_M}=X(1-X)^{i-1},
\end{equation}
showing that the system can be regarded as a polydisperse mixture of monomers, dimers, trimers, etc., with decreasing mole fractions $f_1>f_2>f_3>\cdots$.

The average number of segments in the chain molecules, $\langle i\rangle=\sum_{i=1}^{\infty} i f_i$, follows directly from Eq.~\eqref{eq:clusterSizeDistribution} as
\begin{equation}
\label{eq:meanClusterSize}
\langle i\rangle=\frac{1}{X}.
\end{equation}
Thus, the average physical length of the chains is $\sigma/X$ in the sticky limit ($\delta\to0$), while for finite-range interactions, it lies between $\sigma/X$ and $(\sigma+\delta)/X$.

Equations~\eqref{eq:monomerDensity} and \eqref{eq:meanClusterSize} highlight that $X$ controls both the density of chain molecules and the degree of polymerization, thereby acting as the key thermodynamic parameter of the system.

It is also instructive to establish a connection between the equation of state of the associating system, Eq.~\eqref{eq:pressureFinite}, and that of an additive hard-rod mixture with number density $\rho_M$,
\begin{equation}
\label{eq:mixturePressure}
\bp_M=\frac{\rho_M}{1-\rho_M\sum_{i=1}^nf_i\sigma_i} ,
\end{equation}
where $n$ is the number of components, and $f_i$ and $\sigma_i$ denote the mole fraction and length of species $i$, respectively. In the limit of infinitely many components, identifying $f_i$ with Eq.~\eqref{eq:clusterSizeDistribution} yields
\begin{equation}
\label{eq:pressureFromDistribution}
\bp_M=\frac{\rho X}{1-\rho X^2\sum_{i=1}^{\infty}(1-X)^{i-1}\sigma_i},
\end{equation}
where we have also identified the number density of the mixture with that of chain molecules.

Assuming $\sigma_i=\sigma+(i-1)x_m$, where $x_m$ is an effective value of the average distance between consecutive monomers within a chain, one obtains
\begin{equation}
\label{eq:pressureMeanBondDistance}
\bp_M=\frac{\rho X}{1-\rho\left[x_m-(x_m-\sigma)X\right]}.
\end{equation}
Identifying $p_M=p$ leads to
\beq
\label{xm}
x_m=\sigma+\frac{1}{1-X}\left(\frac{1-\rho\sigma}{\rho}-\frac{X}{\bp}\right).
\eeq
Using Eqs.~\eqref{eq:lawMassActionFinite} and \eqref{eq:pressureFinite}, this expression provides the TPT1 prediction for $x_m$ as a function of $\rho$, $\tau$, and $\delta$. In the low-density limit, $\lim_{\rho\sigma\to 0} x_m=\sigma+\delta/2$, whereas $\lim_{\rho\sigma\to 1} x_m=\sigma$ in the close-packing limit. Moreover, this prediction for $x_m$ exhibits a nonmonotonic dependence on $\rho$, with a maximum at intermediate densities. As expected, $x_m=\sigma$ in the sticky limit.

While the standard TPT1  is exact for 1D sticky
particles,\cite{GV25} it is only approximate for finite-range interactions.
This motivates a comparison with the known exact solution of the equivalent SW fluid, summarized in
Sec.~\ref{sec:ND_theory}, from which both thermodynamic and structural
properties can be obtained directly from the nearest-neighbor distribution
function.

	\section{Exact solution}\label{sec:ND_theory}

As discussed above, the present model is mathematically equivalent to the exactly solvable 1D SW fluid. Therefore, the exact equation of state, nearest-neighbor distributions, and pair correlation functions presented below follow from previously established results.\cite{S16} The novel aspect of the present treatment is their reformulation in terms of association-theory variables and the derivation of exact relations involving bonding quantities, such as the fraction of unbonded sites, the generalized law of mass action, the bonding free energy, the chemical equilibrium constant, and chain statistics.

\subsection{General framework}

The exact structural and thermodynamic properties of 1D systems with nearest-neighbor interactions can be obtained in the isothermal--isobaric ensemble by exploiting the nearest-neighbor distribution function and constructing higher-order distributions through iterated convolutions.\cite{S16}

A central quantity in this approach is the Laplace transform of the Boltzmann factor $e^{-\beta u(x)}$, defined as
\bal
\label{eq:omegaTransform}
\Omega(s)=&\int_{0}^{\infty}dx\,e^{-\beta u(x)}e^{-sx}\nn
=&\frac{e^{-s\sigma}}{s}\left(1+\frac{1-e^{-s\delta}}{\tau\delta/\sigma}\right) ,
\eal
where $s$ is, in general, a complex variable.

The probability density of finding two neighboring particles separated by a distance $x$ is given by\cite{S16}
\begin{equation}
\label{eq:neighborDistanceDistribution}
f^{(1)}(x)=\frac{e^{-\beta u(x)}e^{-\bp x}}{\Omega(\bp)} ,
\end{equation}
where $\Omega(\bp)$ acts as a normalization factor. Moreover, it is directly related to the Gibbs free energy via $\beta G/N=-\ln\Omega(\bp)$. Using the thermodynamic identity $1/\rho=\partial(\beta G)/\partial(\bp)$, one obtains the exact equation of state
\begin{equation}
\label{eq:exactEquationOfState}
\frac{1}{\rho}=\sigma+\frac{1}{\bp}-\frac{\delta e^{-\bp\delta}}{1+\tau\delta/\sigma-e^{-\bp\delta}} .
\end{equation}
Equivalently, this result can be expressed as $1/\rho=\langle x\rangle^{(1)}$, where
\begin{equation}
\label{eq:meanNeighborDistance}
\langle x\rangle^{(1)}=\int_{0}^{\infty}dx\,f^{(1)}(x)x
\end{equation}
is the mean nearest-neighbor distance.
Note that, except in the sticky limit, the pressure cannot be expressed as an explicit function of density from Eq.~\eqref{eq:exactEquationOfState} in terms of elementary functions.

For later convenience, we introduce the reduced variables $\delta^*=\delta/\sigma$,  $\rho^*=\rho\sigma$, and $p^*=\bp\sigma$, which will be used throughout the remainder of the paper.

\subsection{Relation between $X$, $\rho$, and $p$}

The nearest-neighbor distribution function allows one to determine the exact fraction of unbonded sites,
\bal
\label{eq:freeSiteFractionPressure}
X=&\int_{\sigma+\delta}^{\infty}dx\,f^{(1)}(x)=\frac{e^{-(\sigma+\delta)\bp}}{\bp\Omega(\bp)}\nn
=&e^{-p^*\delta^*}\left(1+\frac{1-e^{-p^*\delta^*}}{\tau\delta^*}\right)^{-1}.
\eal
This expression satisfies the expected limits $\lim_{p^*\to 0}X=1$ and $\lim_{p^*\to\infty}X=0$. It can be inverted to yield
\begin{equation}
\label{eq:pressureFromFreeFraction}
p^*=\frac{\ln\alpha(X)}{\delta^*},\quad \alpha(X)\equiv \frac{X+\tau\delta^*}{X(1+\tau\delta^*)}.
\end{equation}
Substitution into Eq.~\eqref{eq:exactEquationOfState} leads to
\begin{equation}
\label{eq:generalizedMassAction}
\frac{1}{\rho^*}=1+\frac{\delta^*}{\ln\alpha(X)}-\frac{X}{\tau} .
\end{equation}

Equation~\eqref{eq:generalizedMassAction} provides an exact analytical relation between $\rho$ and $X$ and may, thus, be interpreted as a generalized law of mass action. Comparison with Eq.~\eqref{eq:lawMassActionFinite} shows that TPT1 does not reproduce this relation for finite-range interactions. However, in the sticky limit ($\delta\to 0$), Eq.~\eqref{eq:generalizedMassAction} reduces to Eq.~\eqref{eq:lawMassActionSticky}, demonstrating that TPT1 becomes exact in that case. Consistently, Eq.~\eqref{eq:freeSiteFractionPressure} yields
\begin{equation}
\label{eq:stickyFreeFractionPressure}
X=\frac{1}{1+p^*/\tau}\quad \text{(sticky limit)},
\end{equation}
in agreement with the combination of Eqs.~\eqref{eq:lawMassActionSticky} and \eqref{eq:pressureSticky}.

The origin of the deviation of TPT1 is that, for finite-range attractions, bonding is controlled by the pressure-dependent distribution of nearest-neighbor separations within the attractive shell, rather than solely by contact configurations. This effect disappears in the sticky limit, where the attractive range collapses to contact and the standard TPT1 law becomes exact.

\begin{figure}
\includegraphics[width=0.8\columnwidth]{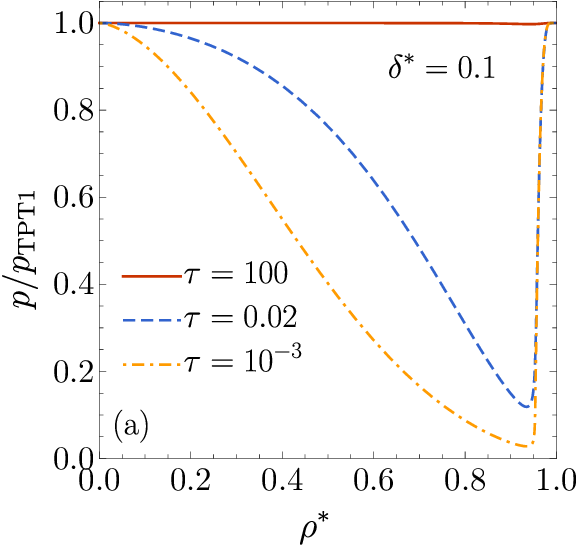}\\
\includegraphics[width=0.8\columnwidth]{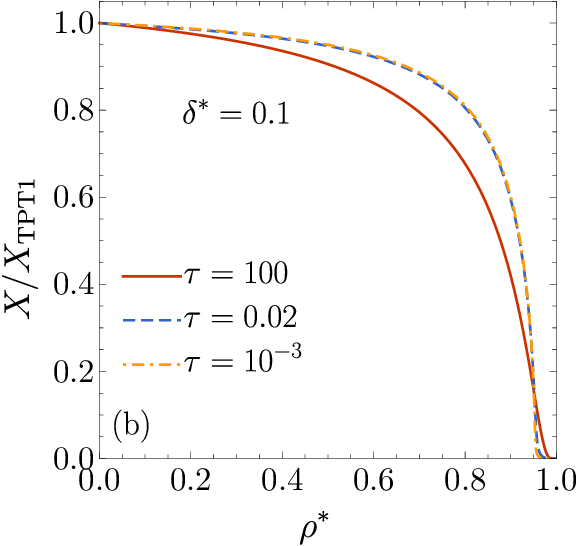}
\caption{Density dependence of the ratios (a) $p/p_{\text{TPT1}}$ and (b) $X/X_{\text{TPT1}}$ for divalent patchy hard rods with $\delta^*=0.1$ and  $\tau=100$, $0.02$, and  $10^{-3}$.}
\label{fig:TPT1}
\end{figure}

The ratio $p/p_{\text{TPT1}}$ between the exact pressure and  that predicted by TPT1  [see Eqs.~\eqref{eq:pressureFinite} and \eqref{eq:exactEquationOfState}] is illustrated in Fig.~\ref{fig:TPT1}(a) for $\delta^*=0.1$ and three values of $\tau$: $100$, $0.02$, and $10^{-3}$. As stickiness increases (i.e., $\tau$ decreases), the limitations of the TPT1 predictions become more and more apparent.   A similar comparison, but in terms of the ratio $X/X_{\text{TPT1}}$, is presented in Fig.~\ref{fig:TPT1}(b). In this case, essentially due to the unphysical TPT1 prediction given by Eq.~\eqref{10}, the disagreement between $X$ and $X_{\text{TPT1}}$ is quite clear, even at high $\tau$. Interestingly, for sufficiently small $\tau$, the ratio $X/X_{\text{TPT1}}$ is practically independent of $\tau$.

\subsection{Chemical equilibrium constant of the chain formation}\label{sec:CE_theory}

The association between sites induces a reversible polymerization process, leading to a polydisperse fluid of chain-like aggregates whose size distribution is governed by Eqs.~\eqref{eq:clusterSizeDistribution} and \eqref{eq:freeSiteFractionPressure}. The bonding between an $i$-mer and a $j$-mer can be formally interpreted as a chemical reaction of the type
\begin{equation}
\label{eq:chainAssociationReaction}
A_i + A_j \overset{K}{\longleftrightarrow} A_{i+j},
\end{equation}
where $K$ denotes the equilibrium constant. The corresponding law of mass action yields
\begin{equation}
\label{eq:equilibriumConstant}
K=\frac{\rho_{i+j}}{\rho_i\rho_j}=\frac{1-X}{X^2\rho},
\end{equation}
where Eq.~\eqref{eq:clusterDensity} has been used.

In the isothermal--isobaric ensemble, combining Eqs.~\eqref{eq:exactEquationOfState} and \eqref{eq:freeSiteFractionPressure} leads to
\bal
\label{eq:equilibriumConstantFinite}
K=&\sigma\frac{1+\tau\delta^*}{(\tau\delta^*)^2}\left(e^{p^*\delta^*}-1\right)\bigg\{\left(1+\frac{1}{p^*}\right)\nn
&\times\left[(1+\tau\delta^*)e^{p^*\delta^*}-1\right]-\delta^*\bigg\} .
\eal
In the sticky limit ($\delta\to 0$), this expression reduces to
\bal
K=&\sigma\frac{\tau+p^*(\tau+p^*)}{\tau^2}\nn
=&\frac{\sigma}{\tau(1-\rho^*)} \quad \text{(sticky limit)},
\eal
where, in the last step, Eq.~\eqref{EOS-SHS} has been used.

\subsection{Helmholtz free energy}\label{sec:FE_theory}

We now derive the exact association contribution to the Helmholtz free energy as a function of the fraction of unbonded sites, $X$.

Starting from the thermodynamic relation $\bp=\rho^2\partial(\beta F/N)/\partial\rho$ and using Eqs.~\eqref{eq:freeEnergySplit}--\eqref{eq:excessFreeEnergy}, one obtains
\beq
\label{33}
\frac{\partial(\beta F_b/N)}{\partial X}=\left(\frac{\rho^*}{1-\rho^*}-{p^*}\right)\frac{\partial{\rho^*}^{-1}}{\partial X}.
\eeq
Combining Eqs.~\eqref{eq:pressureFromFreeFraction} and \eqref{eq:generalizedMassAction}, we find
\beq
\label{34}
\frac{\rho^*}{1-\rho^*}-{p^*}=\frac{(X/\tau\delta^*)\left[\ln\alpha(X)\right]^2}{\delta^*-(X/\tau)\ln\alpha(X)}.
\eeq
Likewise, differentiation of Eq.~\eqref{eq:generalizedMassAction} yields
\beq
\label{35}
\frac{\partial{\rho^*}^{-1}}{\partial X}=\frac{\tau{\delta^*}^2}{X(X+\tau\delta^*)\left[\ln\alpha(X)\right]^2}-\frac{1}{\tau}.
\eeq
Substituting Eqs.~\eqref{34} and \eqref{35} into Eq.~\eqref{33}, one obtains
\bal
\label{36}
\frac{\partial(\beta F_b/N)}{\partial X}=&\frac{1}{1-(X/\tau\delta^*)\ln\alpha(X)}
\Bigg\{\frac{1}{X+\tau\delta^*}
\nn&
-X\left[\frac{\ln\alpha(X)}{\tau\delta^*}\right]^2\Bigg\}.
\eal

Although the right-hand side of Eq.~\eqref{36} has a nontrivial dependence on $X$, it can be integrated analytically. Imposing the boundary condition $F_b=0$ at $X=1$, the exact bonding free energy is obtained as
\beq
\label{eq:exactBondFreeEnergy}
\frac{\beta F_b}{N}=\ln\left[X\alpha(X) -\frac{X^2\alpha(X)\ln\alpha(X)}{\tau\delta^*}\right]
+\frac{X\ln\alpha(X)}{\tau\delta^*} .
\eeq
This result is significantly more involved than the TPT1 expression $\beta   F_b/N=2\ln X - X + 1$ [cf.~Eq.~\eqref{eq:totalFreeEnergy}]. Nevertheless, both expressions coincide in the sticky limit $\delta\to 0$, confirming once more that TPT1 becomes exact in that limit.

\subsection{Bonded distances and limitations of the mixture mapping}

The exact nearest-neighbor distribution also allows one to determine the average distance between consecutive bonded monomers,
\bal
\label{eq:bondedDistanceDefinition}
x_m=&\frac{\int_{\sigma}^{\sigma+\delta}dx\,f^{(1)}(x)x}{\int_{\sigma}^{\sigma+\delta}dx\,f^{(1)}(x)} \nn
=&\sigma\left(1+\frac{1}{p^*}-\frac{\delta^* e^{-p^*\delta^*}}{1-e^{-p^*\delta^*}}\right).
\eal

This expression satisfies the limits $\lim_{p^*\to 0}x_m=\sigma+\delta/2$ and $\lim_{p^*\to\infty}x_m=\sigma$, in agreement with the corresponding limits obtained within TPT1. However, the TPT1 prediction for $x_m$ deviates from the exact result at finite densities, even at a qualitative level.

Furthermore, inserting the exact $x_m$ and $X$ into Eq.~\eqref{eq:pressureFromDistribution} shows that, in general, $p_M\neq p$, particularly near close packing. This demonstrates that the mapping onto an effective hard-rod mixture breaks down for finite-range interactions, although it becomes exact in the sticky limit.

\subsection{Pair correlations and asymptotic behavior}
\label{sec4F}

The $\ell$th-neighbor distribution function, $f^{(\ell)}(x)$, follows from the recursive convolution relation,\cite{S16}
\beq
\label{convol}
f^{(\ell)}(x)=\int_0^x dx'\, f^{(1)}(x')f^{(\ell-1)}(x-x').
\eeq
The pair correlation function is then given by
\beq
g(x)=\frac{1}{\rho}\sum_{\ell=1}^\infty f^{(\ell)}(x).
\eeq
In Laplace space, one obtains
\bal
\label{eq:laplaceCorrelation}
\widetilde{G}(s)\equiv &\int_0^{\infty} dx\, g(x)e^{-sx}\nn
=&\frac{1}{\rho}\frac{\Omega(s+\bp)}{\Omega(\beta p)-\Omega(s+\bp)} .
\eal

The poles ${s_i}$ of $\widetilde{G}(s)$ are determined by the condition $\Omega(s_i+\bp)=\Omega(\bp)$. In terms of these poles, the pair correlation function can be expressed as
\begin{equation}
\label{eq:rdfResidueExpansion}
g(x)=1+\sum_{s_i}e^{s_i x}\operatorname{Res}\left[\widetilde{G}(s_i)\right],
\end{equation}
where $\operatorname{Res}\left[\widetilde{G}(s_i)\right]$ denotes the corresponding residues.

The asymptotic decay of $g(x)$ is governed by the pole with the smallest (in magnitude) negative real part, $s_{\min}=-\kappa\pm \imath \omega$. If $\omega\neq 0$, the decay is oscillatory, $g(x)-1\sim e^{-\kappa x}\cos(\omega x+\delta)$, whereas for $\omega=0$, it is monotonic, $g(x)-1\sim e^{-\kappa x}$. In both cases, the correlation length is $\xi=1/\kappa$.

Oscillatory decay is associated with the dominance of excluded-volume effects and typically occurs at high densities or temperatures, while monotonic decay reflects the prevalence of attractive interactions at low densities or temperatures. The boundary between these regimes defines the FW line in the temperature-density plane.\cite{FW69} In the monotonic regime, the correlation length exhibits a maximum at fixed temperature, defining the Widom line.\cite{XKBCPSS05}

Beyond the FW line, the behavior of $\xi$ changes qualitatively: it displays a weak maximum followed by a minimum and then a rapid increase. To distinguish this locus from the true Widom line, we introduce the term ECO line. Although not a Widom line in the strict sense, the ECO line provides a useful characterization of structural changes in the oscillatory regime.

\section{Finite-range SW sites vs zero-range sticky sites}\label{sec:SW_SHS}

In this section we analyze how the finite range of the patch--patch interaction modifies the thermodynamic and structural properties of the system relative to the zero-range sticky limit.

\subsection{Equation of state and pressure ratio}

We begin by considering the ratio $R\equiv p/p_T$ between the pressure of the divalent patchy hard-rod system and that of the reference Tonks hard-rod fluid at the same density.

For zero-range sticky sites ($\delta\to0$), Eqs.~\eqref{eq:tonksPressure} and \eqref{eq:pressureSticky} yield the simple result $R = X$. Thus, the pressure ratio coincides with the fraction of unbonded sites.

For finite-range interactions ($\delta>0$), combining Eqs.~\eqref{eq:tonksPressure} and \eqref{eq:exactEquationOfState} gives
\begin{equation}
\label{eq:pressureRatioAlpha}
R=1-\frac{p^*\delta^* e^{-p^*\delta^*}}{1+\tau\delta^*-e^{-p^*\delta^*}}.
\end{equation}
Unlike the sticky case, this ratio exhibits a nonmonotonic dependence on density, approaching $R=1$ both in the ideal-gas limit ($\bp\to0$, $\rho^*\to0$) and in the close-packing limit ($\bp\to\infty$, $\rho^*\to1$). By contrast, for sticky sites, $R=X$ decreases monotonically from $1$ to $0$ as density increases.
This qualitative difference reflects the noncommutativity of the limits $\delta\to0$ and $\bp\to\infty$,
\beq
\lim_{\delta\to 0}\lim_{\bp\to \infty}R=1\neq \lim_{\bp\to \infty}\lim_{\delta\to 0}R=0.
\eeq
Indeed, Eq.~\eqref{EOS-SHS} shows that, near close packing,
\begin{equation}
\label{eq:stickyPressureScaling}
p^*\approx\sqrt{{\tau}p_T^*},\quad \text{(sticky limit)},
\end{equation}
which implies that zero-range sticky rods do not collapse into a single infinitely long rod at close packing but remain distributed among chains of different lengths according to Eq.~\eqref{eq:clusterSizeDistribution}.

For finite $\delta$, the nonmonotonic behavior of $R$ implies the existence of a minimum value $R_{\min}$ at a density $\rho_{\min}$. Since $\bp$ increases monotonically with $\rho$, the condition for the minimum is $\partial R/\partial(\bp)=0$. This leads to
\begin{subequations}
\beq
\label{eq:pressureMinimumAlpha}
p_{\min}^*=\frac{R_{\min}}{\delta^*},\quad \rho_{\min}^*=\frac{1}{1+\delta^*} ,
\eeq
\begin{equation}
		\label{eq:alphaMinimumCondition}
		e^{-R_{\min}}=(1+\tau\delta^*)(1-R_{\min}).
\end{equation}
\end{subequations}
The solution is
\beq
R_{\min}=1+W_0\left(-\frac{e^{-1}}{1+\tau\delta^*}\right),
\eeq
where $W_0(z)$ is the principal branch of the Lambert function, which is the solution of the transcendental equation $x=z e^{-x}$.
For small $\tau\delta^*$, $R_{\min}\approx\sqrt{2\tau\delta^*}$ and $p_{\min}^*\approx \sqrt{2\tau/\delta^*}$.
Interestingly, $\rho_{\min}^*$ is independent of the inverse stickiness parameter $\tau$.

\begin{figure}
\includegraphics[width=0.8\columnwidth]{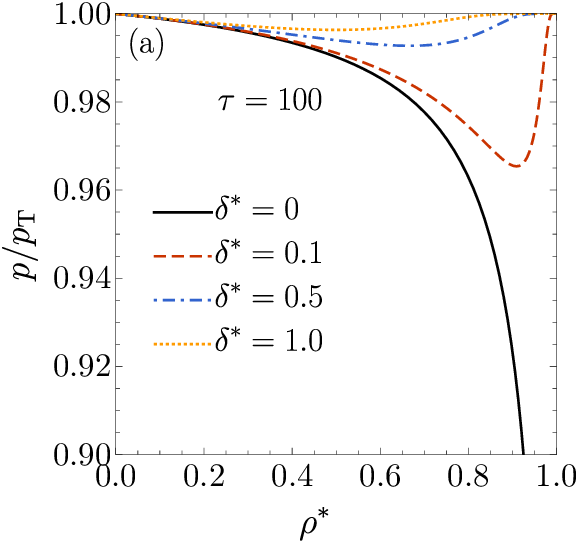}\\
\includegraphics[width=0.8\columnwidth]{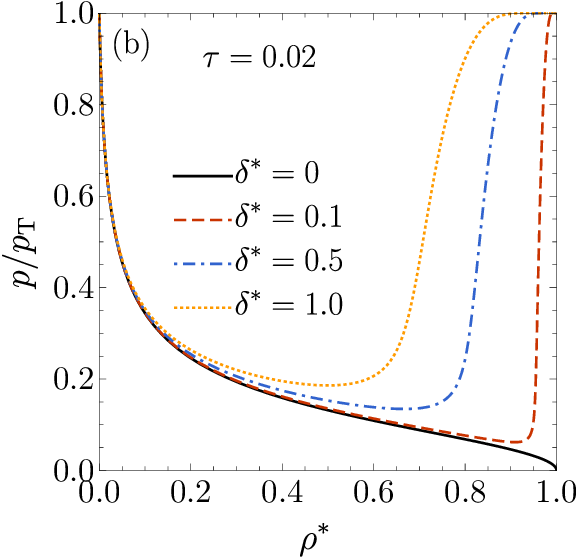}\\
\includegraphics[width=0.8\columnwidth]{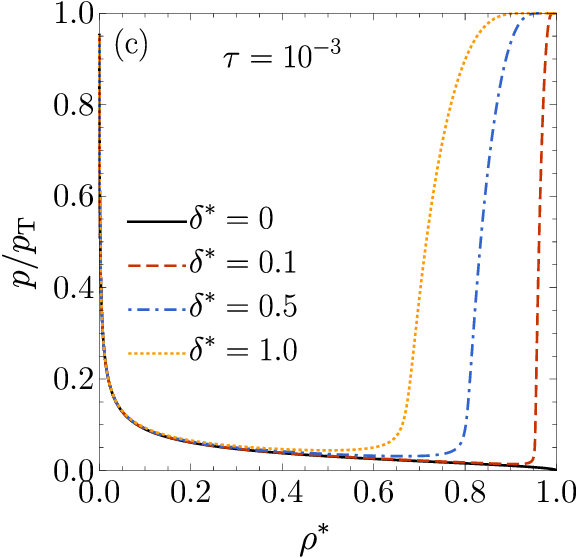}
\caption{Density dependence of the pressure ratio $p/p_T$ for divalent patchy hard rods with (a) $\tau=100$, (b) $\tau=0.02$, and (c) $\tau=10^{-3}$. In each panel, the attractive ranges are $\delta^*=1$, $0.5$, $0.1$, and $0$ (sticky limit).
}
\label{fig:pressure}
\end{figure}

Figure~\ref{fig:pressure} shows the density dependence of $R=p/p_T$ for $\delta^*=1$, $0.5$, $0.1$, and the sticky case $\delta=0$, for three representative values of $\tau$. Panel (a) corresponds to high temperatures (low stickiness), whereas panel (c) corresponds to low temperatures (high stickiness).
The distinction between sticky and finite-range systems becomes especially pronounced near close packing, where $R\to1$ for SW sites but $R\to0$ for sticky ones. At low and intermediate densities, however, the ratio $R$ depends only weakly on $\delta$. This weakly sensitive region widens as either $\delta$ or $\tau$ decreases, extending approximately up to $\rho_{\min}$ for sufficiently small $\tau$.

For $\rho>\rho_{\min}$, the difference becomes significant because neighboring particles are close enough for their attractive ranges to overlap. In the finite-range case, this causes the bonding process to saturate, whereas in the sticky case bonding continues up to close packing.

\subsection{Monomers, terminal rods, and internal rods}

Each particle can exist in three distinct bonding states:
(i) a monomeric state, where both sites are unbonded;
(ii) a terminal state, where only one site is bonded;
(iii) an internal state, where both sites are bonded.
The corresponding average fractions are directly related to the fraction of unbonded sites $X$ through
\begin{equation}
	\label{eq:zeroBondFraction}
	\langle X_0\rangle=X^2 ,
\quad
		\langle X_1\rangle=2X(1-X) ,
\quad
	\langle X_2\rangle=(1-X)^2 .
\end{equation}
These fractions satisfy the exact normalization condition   $\langle X_0\rangle+\langle X_1\rangle+\langle X_2\rangle=1$.

\begin{figure}
	\centering
	\includegraphics[width=0.8\columnwidth]{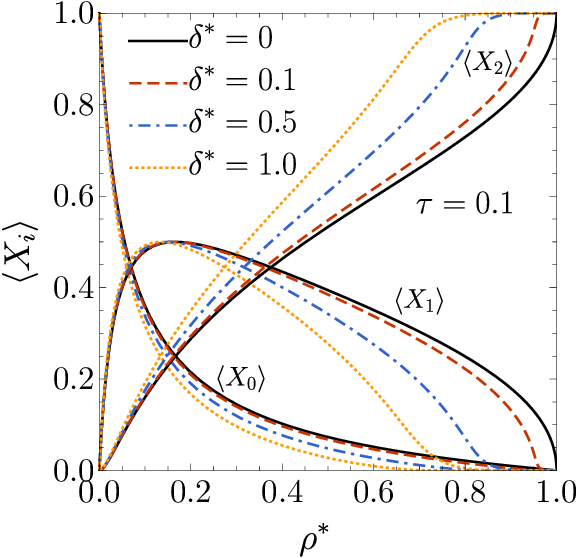}
	\caption{Fraction of free monomers ($\langle X_0\rangle$), terminal rods ($\langle X_1\rangle$), and internal rods ($\langle X_2\rangle$) as  functions of density for $\tau=0.1$ and $\delta^*=1$, $0.5$, $0.1$, and $0$ (sticky limit).}
	\label{fig:Xi}
\end{figure}

Figure~\ref{fig:Xi} illustrates the density dependence of these quantities. In the ideal-gas limit, particles are predominantly monomeric:
$\langle X_0\rangle=1$ and $\langle X_1\rangle=\langle X_2\rangle=0$.
At close packing, particles belong to long chains: $\langle X_0\rangle=\langle X_1\rangle=0$ and $\langle X_2\rangle=1$.
As $\delta$ increases, the approach of $\langle X_2\rangle$ toward unity occurs at lower densities, showing that a larger attractive range favors chain formation.

The fraction of terminal particles reaches a maximum value $\langle X_1\rangle=\frac{1}{2}$, which corresponds to $\langle X_0\rangle=\langle X_2\rangle=\frac{1}{4}$.
Setting $X=\frac{1}{2}$ in Eqs.~\eqref{eq:pressureFromFreeFraction} and \eqref{eq:generalizedMassAction} gives
\begin{subequations}
\begin{equation}
	\label{eq:pressureAtHalfBondFraction}
	p_{\frac{1}{2}}^*=\frac{\ln\frac{1+2\tau\delta^*}{1+\tau\delta^*}}{\delta^*},
\end{equation}
\begin{equation}
	\label{eq:densityAtHalfBondFraction}
	\rho_{\frac{1}{2}}^*=\left(1-\frac{1}{2\tau}+\frac{\delta^*}{\ln\frac{1+2\tau\delta^*}{1+\tau\delta^*}}\right)^{-1}.
\end{equation}
\end{subequations}
In the sticky limit these reduce to $p_{\frac{1}{2}}^*=\tau$ and $\rho_{\frac{1}{2}}^*=2\tau/(1+2\tau)$.
For sufficiently small $\tau$, $\rho_{\frac{1}{2}}^*$ depends only weakly on $\delta$, as shown in Fig.~\ref{fig:Xi}. Moreover, decreasing $\tau$ shifts the maximum of $\langle X_1\rangle$ toward lower densities, thereby favoring chain formation.

Since $\rho_{\frac{1}{2}}^*$ increases monotonically with $\tau$, its upper bound is obtained in the limit $\tau\to\infty$, $\rho_{\frac{1}{2}}^*<1/(1+\delta^*/\ln 2)<\rho_{\min}^*=1/(1+\delta^*)$.

\subsection{Correlation length and structural lines}

We next analyze the correlation length $\xi$ together with the FW, Widom, and ECO lines introduced in Sec.~\ref{sec4F}.

\begin{figure}
	\includegraphics[width=0.8\columnwidth]{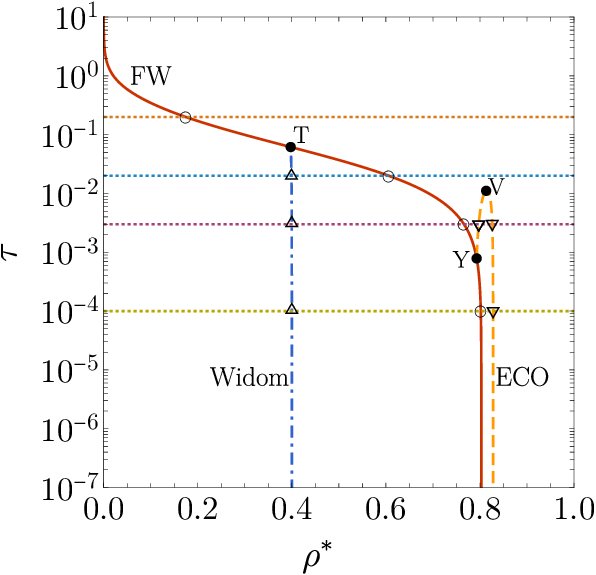}
	\caption{Structural diagram in the $\tau$--$\rho^*$ plane for divalent patchy hard rods with $\delta^*=0.5$. The FW line (solid curve) separates the regions of monotonic and oscillatory asymptotic decay of the pair correlation function. Below the FW line, the Widom line (dash-dotted curve) is defined as the locus of absolute maxima of the correlation length $\xi$ in the monotonic regime. In the oscillatory regime, $\xi$ may exhibit a weak local maximum followed by a local minimum; the locus of these extrema defines the ECO line (dashed curve).  The horizontal dotted lines correspond to $\tau=0.2$, $0.02$, $3\times10^{-3}$, and $10^{-4}$. The circles, upward triangles, and downward triangles indicate the intersections with the FW, Widom, and ECO lines, respectively.}
	\label{fig:fancylines}
\end{figure}

At fixed interaction range $\delta$, the FW line divides the $\tau$--$\rho^*$ plane into two regions: above the line, the asymptotic decay of $g(x)-1$ is oscillatory, whereas below it, the decay is monotonic. This line is illustrated in Fig.~\ref{fig:fancylines} for the representative case $\delta^*=0.5$.
The FW line starts at $\rho^*=0$ in the limit $\tau\to\infty$, since the hard-rod fluid exhibits oscillatory decay at all densities, and terminates at
$\rho^*=\rho^*_0\simeq 1/(1+\delta^*/2)=2\rho^*_{\min}/(1+\rho^*_{\min})$
in the opposite limit $\tau\to0$ (see Appendix~\ref{appA}). Therefore, independently of the stickiness, the decay of correlations is always oscillatory for $\rho^*>\rho_0^*$.

Crossing the FW line produces a kink in the correlation length due to the change in the leading pole of $\widetilde{G}(s)$: the dominant pole changes from a real pole,
$s_{\min}=-\kappa$, to a complex-conjugate pair,
$s_{\min}=-\kappa\pm\imath\omega$.

The Widom line has physical meaning only in the monotonic region below the FW line, where the correlation length reaches an absolute maximum. This line is nearly vertical in the $\tau$--$\rho^*$ plane and approaches
$\rho_c^*=1/(2+\delta^*)\simeq \rho_0^*/2$ as $\tau\to0$.

In higher-dimensional systems, the Widom line is commonly regarded as an extension of the vapor--liquid coexistence region beyond the critical point and terminates at the critical point, where $\xi \to \infty$.\cite{FW69,SHER19} Since no true phase transition can occur in the present 1D system, the Widom line should instead be interpreted as a precursor of the vapor--liquid transition that would occur in higher dimensions. From this perspective, the point $(\rho^*, \tau) = (\rho_c^*, 0)$, where $\xi \to \infty$ (see Appendix \ref{appA}), may be regarded as a putative 1D critical point.

Beyond the FW line, in the oscillatory regime, the damping constant still defines an effective correlation length, but its density dependence becomes more complex. In particular, $\xi$ may exhibit a weak local maximum followed by a local minimum before increasing rapidly again. To characterize this behavior, we introduce the ECO line, which joins the loci of these extrema.

The left branch of the ECO line corresponds to the local maximum of $\xi$, while the right branch corresponds to the local minimum. Although the ECO line is not a Widom line in the strict sense, it provides a natural structural boundary within the oscillatory regime. To the best of our knowledge, this line has not been identified previously.
It is possible that an analogous feature also exists in higher-dimensional systems, although likely hidden inside metastable regions associated with solid--solid transitions. In that context, the ECO line might be related to the isostructural solid--solid transition observed in SW systems.\cite{BHF94,LNL94}

For later reference, we introduce labels for the three characteristic points where these structural lines intersect or merge. The intersection of the FW and Widom lines is denoted the T-point, since the two curves form a T-like structure. Similarly, the intersection of the FW and ECO lines is called the Y-point, while the vertex where the two ECO branches merge is referred to as the V-point. These special points provide a convenient framework for organizing the topology of the structural map in the $(\rho^*,\tau)$ plane.

The density dependence of $\xi$ changes qualitatively depending on the value of $\tau$.
If $\tau>\tau_{\mathrm{T}}$, the correlation length increases monotonically with density and presents only a kink at the FW point. This behavior is represented by the case $\tau=0.2$ in Fig.~\ref{fig:fancylines}.
For $\tau_{\mathrm{T}}<\tau<\tau_{\mathrm{V}}$, represented by $\tau=0.02$, $\xi$ first increases with density, reaches an absolute maximum upon crossing the Widom line, and then increases monotonically again after crossing the FW line.
The behavior becomes richer for $\tau_{\mathrm{V}}<\tau<\tau_{\mathrm{Y}}$, illustrated by $\tau=3\times10^{-3}$. In this case, $\xi$ first reaches a maximum at the Widom line, then increases after crossing the FW line, reaches a local maximum followed by a local minimum within the oscillatory regime, and finally increases monotonically again.
Finally, for $\tau<\tau_{\mathrm{Y}}$, as in the case $\tau=10^{-4}$, the correlation length continues decreasing after crossing the FW line, reaches a local minimum, and only then starts its final monotonic growth.

We have verified that these structural crossovers are not accompanied by any anomalous behavior of local bonding quantities, such as the fraction of unbonded sites, the mean chain length, or higher moments of the chain-size distribution, all of which vary smoothly across the ECO line. This indicates that the ECO line is a purely structural feature associated with the long-range asymptotic behavior of positional correlations.

\begin{figure}
	\centering
	\includegraphics[width=0.8\columnwidth]{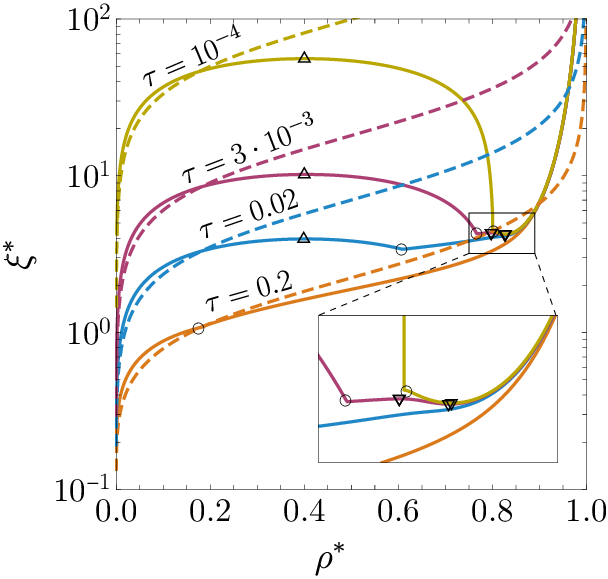}
	\caption{Reduced correlation length $\xi^*$ as a function of reduced density $\rho^*$ for $\tau=0.2$, $0.02$, $3\times10^{-3}$, and $10^{-4}$. The solid lines correspond to finite-range SW sites with $\delta^*=0.5$, while dashed lines correspond to the sticky limit ($\delta\to0$).  The inset magnifies the region $0.7<\rho^*<0.9$. The circles, upward triangles, and downward triangles indicate the intersections with the FW, Widom, and ECO lines, respectively.}
	\label{fig:correlation}
\end{figure}

Figure~\ref{fig:correlation} shows the density dependence of the reduced correlation length $\xi^*=\xi/\sigma$ for $\delta^*=0.5$ and the same four values of $\tau$ as those indicated in Fig.~\ref{fig:fancylines}. For comparison, the corresponding curves for the sticky limit ($\delta\to0$) are also included.
A major difference appears in the sticky case: unlike finite-range systems with $\tau<\tau_{\mathrm{T}}$, the correlation length increases monotonically with density for all values of $\tau$. In fact, as shown in Appendix~\ref{appB},
\beq
\tau_{\mathrm{T}}
\approx
\frac{\pi^4}{36}{\delta^*}^4,
\quad
\delta^*\ll 1.
\eeq
Thus, as $\delta^*\to0$, both the FW and Widom lines collapse toward the axis $\tau=0$, implying that in the sticky limit, the decay of correlations is always oscillatory and $\xi$ always increases monotonically with density.

For finite $\delta$, once the ECO line has been crossed, $\xi$ increases rapidly with density. Interestingly, the asymptotic high-pressure behavior of $\xi$ is qualitatively different for finite- and zero-range patches (see Appendix~\ref{appC}):
\beq
\label{57}
\xi\sim
\begin{cases}
p^2,& \delta\neq 0,\\
p^3,&\delta=0.
\end{cases}
\eeq
Therefore, the high-pressure correlation length of hard rods with finite-range patches behaves as in the Tonks gas, which also follows the scaling law $\xi\sim p^2$.\cite{MS26} This is fully consistent with Fig.~\ref{fig:pressure}, where we found that $p/p_T\to1$ near close packing.
By contrast, hard rods with zero-range sticky sites behave differently both in their equation of state ($p/p_T\to0$) and in their correlation length ($\xi\sim p^3$). This reflects the fact that the bonding process remains incomplete until close packing, leading to stronger positional correlations.

\begin{figure}
	\includegraphics[width=0.8\columnwidth]{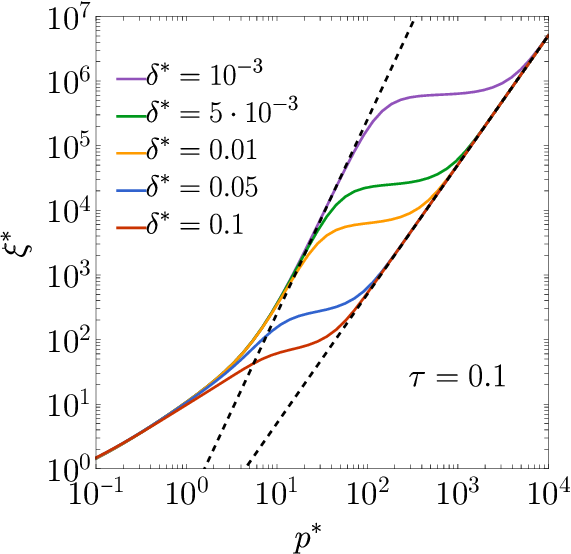}
	\caption{Log--log plot of the reduced correlation length $\xi^*$ vs the reduced pressure $p^*$ for $\tau=0.1$ and, from bottom to top, $\delta^*=0.1$, $0.05$, $0.01$, $5\times10^{-3}$, and $10^{-3}$. The left and right dashed lines represent the asymptotic behaviors $\xi^*= {p^*}^3/4\pi^2\tau$ and $\xi^*={p^*}^2/2\pi^2$, respectively.}
	\label{fig:kappa}
\end{figure}

As shown in Appendix~\ref{appC}, if $\delta^*\ll 1$, an intermediate high-pressure regime appears in which $\xi\sim p^3$ if $p^*\delta^*\ll 1$, followed by a crossover to the ultimate asymptotic regime $\xi\sim p^2$ for $p^*\delta^*\gg 1$. The crossover occurs around $p^*\sim 1/\delta^*$.
This behavior is confirmed in Fig.~\ref{fig:kappa} for $\tau=0.1$ and $\delta^*=0.1$, $0.05$, $0.01$, $5\times10^{-3}$, and $10^{-3}$. As $\delta^*$ decreases, the crossover shifts to higher pressures and the sticky-like regime becomes progressively more extended.

\section{Conclusions}\label{sec:conclusions}

In this work we have investigated the exact thermodynamic and structural properties of divalent patchy hard rods confined to a strictly 1D channel. We have used this model both as an exact benchmark for Wertheim's association theory and as a framework for uncovering novel structural phenomena induced by finite-range bonding.

The bonding sites are modeled as attractive SW patches located at the tips of the hard rods. The sticky limit is recovered by taking the well depth to infinity ($\epsilon\to\infty$) and the well range to zero ($\delta\to0$), while keeping the stickiness parameter $\tau^{-1}=(e^{\beta\epsilon}-1)\delta/\sigma$ finite. Throughout this work, we have used the inverse stickiness parameter $\tau$ instead of temperature as the relevant thermodynamic variable, since it provides a unified description of finite-range SW systems and their zero-range sticky limit.

Our results show that TPT1 is approximate for finite-range SW sites ($\delta>0$), but it becomes exact in the zero-range sticky limit ($\delta\to0$).
We have shown that the exact solution of the equivalent 1D  SW fluid can be naturally recast in the language of association theories. This reformulation provides exact expressions for bonding quantities, including a generalized law of mass action relating $\rho$ and $X$ [Eq.~\eqref{eq:generalizedMassAction}] and an exact bonding free-energy contribution [Eq.~\eqref{eq:exactBondFreeEnergy}].
This generalized law of mass action highlights the limitations of the standard finite-range expression [Eq.~\eqref{eq:lawMassActionFinite}], while correctly recovering the sticky result [Eq.~\eqref{eq:lawMassActionSticky}] in the limit $\delta\to0$.

A key result is the derivation of an exact bonding free-energy contribution as an explicit function of $X$, which preserves the formal simplicity of the TPT1 framework. Together with the generalized law of mass action, this expression provides an exact alternative to the standard TPT1 bonding term and mass-action equation for 1D systems with finite-range patches.
Extending this exact reformulation to more general bonding schemes, such as non-SW interactions\cite{MS19} or Janus-like particles,\cite{MS20,FMS21} and especially to higher-dimensional systems, remains a challenging problem. Nevertheless, the exact 1D relation between density and bonding variables suggests that systematic corrections to TPT1 might be constructed through expansions around the sticky limit or the low-density regime, thereby incorporating finite bond-volume effects beyond the standard contact approximation.

While TPT1 is naturally formulated in the canonical ensemble, the exact solution is obtained in the isothermal--isobaric ensemble, where the density can be analytically expressed as a function of pressure, although the inverse relation does not admit a closed form except in the sticky limit. Conversely, the exact solution is restricted to strictly 1D geometries, whereas the standard TPT1 formalism remains readily applicable to higher-dimensional systems.

The exact solution also provides direct access to structural properties, a capability that is not inherently available within standard association theories. Although true thermodynamic phase transitions are absent in 1D fluids of divalent patchy hard rods,\cite{CS04} the pair correlation function $g(x)$ exhibits a rich structural behavior. In particular, the asymptotic decay of $g(x)-1$ changes from monotonic to oscillatory as density increases, the FW line separating both regimes. In the monotonic regime, the correlation length  displays an absolute maximum, giving rise to the Widom line, while in the oscillatory regime, it may exhibit both a local maximum and a local minimum, giving rise to the ECO line. These features define a remarkably rich structural map in the $(\tau,\rho)$ plane, but they disappear in the sticky limit, where the system remains in the oscillatory regime for any $\tau>0$.

A particularly important result is the identification of the ECO line, a structural locus not previously reported in other 1D fluids.\cite{MRYSH24,MYSH25} This line appears at high densities within the oscillatory regime and connects the extrema of the correlation length beyond the FW crossover. Because of its connection with strong positional ordering, we speculate that the ECO line may represent a structural precursor of freezing or of isostructural solid--solid transitions in two- and three-dimensional systems. Whether an analogous feature exists in other 1D models or in higher-dimensional fluids remains an open and interesting question.
In this sense, the ECO line reveals an additional layer of structural complexity hidden in apparently simple systems confined to restricted geometries.\cite{MS24}

Finally, we identified a qualitative change in the high-pressure behavior of the correlation length. For finite-range SW sites, the asymptotic behavior is $\xi\sim p^2$, identical to that of ordinary hard rods, while for zero-range sticky interactions, the stronger divergence $\xi\sim p^3$ is found. If the attractive range is very small but finite, an intermediate sticky-like regime with $\xi\sim p^3$ appears at high pressure under the condition $\bp\delta\ll1$, followed by a crossover to the true asymptotic finite-range behavior $\xi\sim p^2$ when $\bp\delta\gg1$. This reflects the noncommutativity of the limits $\delta\to0$ and $p\to\infty$ and further emphasizes that finite-range and sticky interactions, although closely related, are not thermodynamically equivalent.

Overall, our results show that even the simplest 1D patchy fluid exhibits a remarkably rich interplay between association, structure, and asymptotic correlations. Beyond providing an exact benchmark for association theories, the present model highlights the nontrivial role of finite bond volumes and offers a useful reference framework for understanding reversible self-assembly in confined geometries and, more broadly, in complex patchy systems.

\begin{acknowledgments}
A.M.M. and A.S. acknowledge financial support from Grant No.~PID2024-156352NB-I00 funded by MCIU/AEI/10.13039/501100011033 and by ERDF/EU, and from Grant No.~GR24022 funded by the Junta de Extremadura (Spain). A.M.M. is grateful to the Spanish Ministerio de Ciencia e Innovaci\'on for a fellowship Grant No.~RE2021-097702.
P.G. and S.V. gratefully acknowledge the financial support of the National Research, Development and Innovation Office--Grants Nos.~2023-1.2.4-TET-2023-00007 and K137720.
\end{acknowledgments}

\section*{AUTHOR DECLARATIONS}

\subsection*{Conflict of Interest}

The authors have no conflicts to disclose.

\subsection*{Author Contributions}
\textbf{Ana M. Montero}:
Conceptualization (equal);
Formal analysis (equal);
Funding acquisition (equal);
Investigation (equal);
Methodology (equal);
Software (equal);
Validation (equal);
Visualization (lead);
Writing -- original draft (supporting);
Writing -- review \& editing (supporting).
\textbf{Andr\'es Santos}:
Conceptualization (equal);
Formal analysis (equal);
Funding acquisition (equal);
Investigation (equal);
Methodology (equal);
Software (equal);
Validation (equal);
Visualization (supporting);
Writing -- original draft (equal);
Writing -- review \& editing (equal).
\textbf{P\'eter Gurin}:
Formal analysis (equal);
Funding acquisition (equal);
Investigation (equal);
Methodology (supporting);
Writing -- original draft (supporting);
Writing -- review \& editing (supporting).
\textbf{Szabolcs Varga}:
Conceptualization (lead);
Formal analysis (equal);
Funding acquisition (equal);
Investigation (equal);
Methodology (lead);
Validation (equal);
Writing -- original draft (lead);
Writing -- review \& editing (equal).

\section*{DATA AVAILABILITY}

The data that support the findings of this study are available from the corresponding author upon
reasonable request.

\appendix
\section{Widom and FW lines in the high-stickiness limit ($\tau\ll 1$)}
\label{appA}

\subsection{Low-temperature end point of the Widom line}

The real poles of $\widetilde{G}(s)$ are determined by the condition
$\Omega(-\kappa+\bp)=\Omega(\bp)$, where $\Omega(s)$ is given by Eq.~\eqref{eq:omegaTransform}. More explicitly,
\beq
\label{3z}
\frac{1+\tau\delta^*-e^{-p^*\delta^*}}{p^*}=\frac{e^{\kappa^*}}{p^*-\kappa^*}\left[1+\tau\delta^*
-e^{-(p^*-\kappa^*)\delta^*}\right],
\eeq
where $\kappa^*\equiv\kappa\sigma$.

The Widom line is determined by the condition $[\partial\kappa/\partial(\bp)]_\tau=0$. Therefore, in addition to the condition $\Omega(-\kappa+\bp)=\Omega(\bp)$, we must also impose
$\Omega'(-\kappa+\bp)=\Omega'(\bp)$,
where $\Omega'(s)\equiv \partial\Omega(s)/\partial s$.
For given $\tau$ and $\delta^*$, the values of $\kappa^*$ and $p^*$ on the Widom line are obtained from the simultaneous solution of Eq.~\eqref{3z} and
\beq
\label{3y}
1-\frac{\kappa^*}{p^*}+\frac{\left(1+\tau\delta^*\right)e^{p^*\delta^*}-1}{\delta^*}\frac{\kappa^*}{{p^*}^2}=e^{\kappa^*(1+\delta^*)}.
\eeq

In the high-stickiness regime ($\tau\ll1$), the solution of Eqs.~\eqref{3z} and \eqref{3y} is
\beq
\label{A3}
\kappa^*\approx 2p^*,\quad p^*\approx\sqrt{\frac{\tau}{1+\delta^*/2}}.
\eeq
Insertion into the exact equation of state, Eq.~\eqref{eq:exactEquationOfState}, yields the low-temperature end point of the Widom line,
\beq
\label{A4}
\rho^*_c=\frac{1}{2+\delta^*}.
\eeq
From Eq.~\eqref{A3}, we observe that $\kappa\to 0$ in the limit $\tau\to 0$. Thus, the correlation length diverges as one approaches the ``critical point'' $(\rho^*,\tau)=(\rho_c^*,0)$.

\subsection{Low-temperature end point of the FW line}

The complex poles satisfy $\Omega(-\kappa\pm\imath\omega+\bp)=\Omega(\bp)$, namely,
\begin{subequations}
\label{3}
\bal
\label{3a}
\frac{1+\tau\delta^*-e^{-p^*\delta^*}}{p^*}=&\frac{e^{\kappa^*}}{p^*-\kappa^*}\Big\{\left(1+\tau\delta^*\right)
\cos\omega^*
\nn&
-e^{-(p^*-\kappa^*)\delta^*}
\cos[\omega^*(1+\delta^*)]\Big\},
\eal
\bal
\label{3b}
\frac{1+\tau\delta^*-e^{-p^*\delta^*}}{p^*}=&-\frac{e^{\kappa^*}}{\omega^*}\Big\{\left(1+\tau\delta^*\right)\sin\omega^*\nn
&-e^{-(p^*-\kappa^*)\delta^*}
\sin[\omega^*(1+\delta^*)]\Big\},
\eal
\end{subequations}
where $\omega^*\equiv\omega\sigma$.

For given $\tau$ and $\delta^*$, the set of Eqs.~\eqref{3z} and \eqref{3} determines the values of $p^*$, $\kappa^*$, and $\omega^*$ on the FW line.
In the high-stickiness limit ($\tau\to0$), one finds  $\omega^*\to 2\pi/(1+\delta^*/2)$, so that $\omega^*(1+\delta^*)\to 4\pi-\omega^*$, and $\kappa^*\to p_0^*$, where $p^*=p_0^*$ is the solution of the transcendental equation
\beq
\label{FWp}
\frac{\pi}{1+\delta^*/2}\left(1-e^{-p_0^*\delta^*}\right)+p_0^* e^{p_0^*}\sin\frac{2\pi}{1+\delta^*/2}=0.
\eeq
The corresponding  density $\rho_0^*$ is obtained by inserting
$p^*=p_0^*$ into Eq.~\eqref{eq:exactEquationOfState} with $\tau=0$.
An excellent approximation follows from a first-order expansion in $\delta^*$:
\beq
\rho_0^*\simeq \frac{1}{1+\delta^*/2}=2\rho_c^*.
\eeq

\section{Collapse of the Widom and FW lines for $\delta^*\ll1$}
\label{appB}

When solving Eqs.~\eqref{3z} and \eqref{3y} for the Widom line, or Eqs.~\eqref{3z} and \eqref{3} for the FW line, one observes that both lines move downward in the $\tau$--$\rho^*$ plane as $\delta^*$ decreases, eventually collapsing onto the axis $\tau=0$ in the sticky limit.
In fact, for sufficiently small $\delta^*$, the lines corresponding to different values of $\delta^*$ collapse onto a common curve in the scaled plane
$\tau/{\delta^*}^4$ vs $\rho^*$.
This indicates that the relevant regime is  $\tau\sim {\delta^*}^4$. In this limit, the exact equation of state, Eq.~\eqref{eq:exactEquationOfState}, gives
\beq
\label{17}
p^*\approx \sqrt{\frac{\tau\rho^*}{1-\rho^*}}\sim{\delta^*}^2.
\eeq

Substituting $\tau\sim{\delta^*}^4$ and $p^*\sim{\delta^*}^2$ into Eqs.~\eqref{3z} and \eqref{3y}, and taking the limit $\delta^*\to0$, one obtains
\beq
\label{B2}
\kappa^*\approx 2p^*,
\quad
p^*\approx \sqrt{\tau},
\quad
\rho^*\approx\frac{1}{2}.
\eeq
This result characterizes the Widom line for $\delta^*\ll1$ and is fully consistent with Eqs.~\eqref{A3} and \eqref{A4}.

Analogously, Eqs.~\eqref{3z} and \eqref{3} yield
\beq
\label{B3}
\omega^*\approx \frac{2\pi}{1+\delta^*/2},\quad \kappa^*\approx p^*+\frac{\tau}{p^*},\quad p^*\approx \frac{\pi^2}{6}{\delta^*}^2.
\eeq
In terms of density,
\beq
\label{B4}
\kappa^*\approx\frac{\pi^2}{6\rho^*}{\delta^*}^2,\quad \tau\approx\frac{\pi^4}{36}\frac{1-\rho^*}{\rho^*}{\delta^*}^4.
\eeq
Equations~\eqref{B3} and \eqref{B4} characterize the FW line in the limit $\delta^*\ll1$.

Particularizing Eq.~\eqref{B4} to $\rho^*=1/2$, we obtain the values at  the T-point,
\beq
\label{B5}
\kappa^*\approx\frac{\pi^2}{3}{\delta^*}^2,\quad \tau\approx\frac{\pi^4}{36}{\delta^*}^4.
\eeq

\section{Correlation length in the high-pressure limit}
\label{appC}

\subsection{Finite-range patches}

In the limit $p^*\gg 1$ at finite $\delta^*$, Eqs.~\eqref{3} reduce to
\beq
\label{4}
1-\frac{\kappa^*}{p^*}=e^{\kappa^*}\cos\omega^*,\quad
-\frac{\omega^*}{p^*}=e^{\kappa^*}\sin\omega^*.
\eeq
These are exactly the same conditions as for the Tonks gas.\cite{MYSH25} The high-pressure solution is, therefore,
\beq
\label{5}
\kappa^*\approx \frac{2\pi^2}{{p^*}^2},\quad\omega^*\approx 2\pi\left(1-\frac{1}{p^*}\right).
\eeq

\subsection{Sticky limit}
Before taking the high-pressure limit, let us first consider $\delta^*\to 0$ in Eqs.~\eqref{3}. This gives
\begin{subequations}
\label{8}
\beq
1+\frac{\tau}{p^*}=\frac{e^{\kappa^*}}{p^*-\kappa^*}\left[\left(\tau+p^*-\kappa^*\right)\cos\omega^*
+\omega^*\sin\omega^*\right],
\eeq
\beq
1+\frac{\tau}{p^*}=-\frac{e^{\kappa^*}}{\omega^*}\left[\left(\tau+p^*-\kappa^*\right)\sin\omega^*
-\omega^*\cos\omega^*\right].
\eeq
\end{subequations}

In the high-pressure limit $p^*\gg 1$, one finds
\beq
\label{9}
\kappa^*\approx \frac{4\pi^2\tau}{{p^*}^3},\quad\omega^*\approx 2\pi\left(1-\frac{\tau}{{p^*}^2}\right).
\eeq

\subsection{Crossover between the $p^{-2}$ and $p^{-3}$ regimes}

The discrepancy between Eqs.~\eqref{5} and \eqref{9} reflects the noncommutativity of the limits $\delta^*\to 0$ and $p^*\to\infty$.
In deriving Eq.~\eqref{5}, we first assumed $e^{-p^*\delta^*}\to 0$ at finite $\delta^*$, and only afterward took $p^*\gg 1$. By contrast, in deriving Eq.~\eqref{9}, we first used
$e^{-p^*\delta^*}\approx 1-p^*\delta^*$ and then considered the limit $p^*\gg 1$.
Therefore, Eq.~\eqref{5} applies when $p^*\delta^*\gg 1$, whereas Eq.~\eqref{9} applies when $p^*\gg 1\gg p^*\delta^*$.

To connect both behaviors, let us assume $\delta^*\ll 1$ and $p^*\gg 1$, while keeping $\pt\equiv p^*\delta^*$ finite.
Introducing $\delta^*=\pt/p^*$ into Eqs.~\eqref{3} and expanding in powers of $1/p^*$, we obtain
\beq
\label{15a}
\kappa^*\approx\frac{2\pi^2}{{p^*}^2}\left(a_0+\frac{a_1}{p^*}\right),\quad \omega^*\approx 2\pi\left(1-\frac{b_1}{p^*}-\frac{b_2}{{p^*}^2}\right),
\eeq
with
\begin{subequations}
\beq
a_0=1-e^{\pt}\theta^2,\quad \theta\equiv \frac{\pt}{e^{\pt}-1},
\eeq
\beq
a_1=\tau e^{\pt}\left(1+e^{\pt}\right)\theta^3-3(1-\theta)\left(1-e^{\pt}\theta^2\right),
\eeq
\beq
b_1=1-\theta,\quad b_2=\tau e^{\pt}\theta^2-\left(1-\theta\right)^2.
\eeq
\end{subequations}

In the limit $\pt\to\infty$ (i.e., $p^*\to\infty$ at fixed $\delta^*$), one has $\theta\to 0$, so that $a_0\to 1$, $a_1\to -3$, $b_1\to 1$, and $b_2\to -1$. Equation~\eqref{5} is then recovered.
By contrast, if $\pt\to 0$ (i.e., $\delta^*\to 0$ at fixed $p^*$), then $\theta\to 1$, so that $a_0\to 0$, $a_1\to 2\tau$, $b_1\to 0$, and $b_2\to \tau$, in agreement with Eq.~\eqref{9}.

Therefore, if $\delta^*$ is small but nonzero, two distinct high-pressure regimes can be identified. First, when $p^*\gg 1$ but $p^*\delta^*\ll 1$, the system behaves effectively as a zero-range sticky fluid, with
$\kappa^*\sim {p^*}^{-3}$. As pressure increases further, the true asymptotic regime $p^*\delta^*\gg 1$ is reached, and the genuine finite-range behavior emerges, with $\kappa^*\sim {p^*}^{-2}$.
The crossover between both regimes takes place in the intermediate region $p^*\sim{\delta^*}^{-1}$.

\section*{References}

\bibliography{C:/AA_D/Dropbox/Mis_Dropcumentos/bib_files/liquid}

\end{document}